\documentclass[preprint,aps,showpacs]{revtex4}
\begin{document} 

\title{Probing Anderson localization of light via decay rate statistics}
\author{ 
F.\,A.\,Pinheiro $^{1}$\footnote{Corresponding author. E-mail: felipe.pinheiro@grenoble.cnrs.fr}, M.\ Rusek $^{2}$, A.\ Orlowski $^{2}$,  
and B.\,A.\,van Tiggelen $^{1}$
}
\affiliation{$^{1}$CNRS/Laboratoire~de~Physique~et~Mod\'elisation~des~Milieux~Condens\'es,
Universit\'e~Joseph~Fourier,~Maison~des~Magist\`eres,\\
B.P. 166~38042~Grenoble~Cedex~9,~France\\
$^{2}$Instytut Fizyki, Polska Akademia Nauk,\\
Aleja Lotnik\'ow 32/46, 02-668 Warszawa, Poland
}
%
\date{\today}
%
%
%
%
\begin{abstract}

We have studied the distribution of resonant widths $P \left(  \Gamma  \right)$ in one-, two- and three dimensional multiple light scattering systems. $P \left(  \Gamma  \right)$ should follow a universal power law $P \left(  \Gamma  \right) \sim \Gamma^{-1}$ in the localized regime as confirmed by extensive numerical calculations. This behavior can be interpreted as an unambiguous signature of exponential Anderson localization of light in open systems.

\end{abstract}
\maketitle
%
%
%
%
%
%
%
%

The research on Anderson localization of light
has been of great interest~\cite{cargese} since it was originally proposed as the
optical counterpart of electronic localization~\cite{john}. 
Localization, as proposed by Anderson, is defined as
an inhibition of wave diffusion in infinite disordered media
due to interference of multiple scattered waves~\cite{anderson}.
A much stronger definition is that the eigenfunctions in an infinite disordered medium are
characterized by an exponential decay in space,
$\left| \psi \left( \mathbf{r} \right)\right| \sim 
\exp  \left(  \left|{\mathbf r}-{\mathbf r}^{\prime}\right|/ \xi \right)$, where
$\xi$ is the localization length.
In finite, {\em open} media, waves can ``leak" through the sample boundaries. 
Anderson localization must thus relate to manifestations
of leakage in observables quantities.
For optical systems, they are typically the
emerging intensity, the total transmission or
the coherent backscattering cone. 
The observation of an exponential scaling of transmission~\cite{azgarcia,nature},
as well as the rounding of the backscattering cone
~\cite{cone}, may not have definitively established localization
since absorption could be responsible for these same effects.

There are several criteria to determine the onset of the localized regime.
The Ioffe-Regel criterion states that, in three dimensions (3D), localization occurs for $k \ell \sim 1$ (with $k$ the light wavenumber inside the medium and $\ell$ the mean free path).
Another approach to define electromagnetic localization is
based on the variance of fluctuations of transmission, even in the presence of absorption~\cite{natureaz}. 
In open systems, the ``eigenstates" are resonances with a finite energy width $\Gamma$ (or, equivalently, with a finite lifetime $t \sim 1/\Gamma$) due to leakage.
The Thouless criterion establishes that localization can be set to occur when
the typical time that an excitation needs to propagate through the entire
system of size $R$, $t_{T} \sim 1/\Gamma_{T} \sim R^{2}/D$ (Thouless time), exceeds the maximal time scale of the system,
$t_{H} \sim 1/\Gamma_{H} \sim 1/\Delta E$ (Heisenberg time)~\cite{thouless}. Here $D$ is the
diffusion constant and $\Delta E$ the mean level spacing.

This Thouless criterion applies to the {\em average} leakage width.
Hence it is reasonable to assume that the {\em statistical properties} of resonance widths
are strongly affected by localization. The aim of the present paper
is to investigate how localization manifests itself in the
distribution of resonance widths $P (\Gamma)$ in 
{\em multiple light scattering in open systems}.
We will show that $P (\Gamma)$ exhibits the
universal power law $P (\Gamma) \sim \Gamma^{-1}$ in 1D, 2D and 3D
{\em optical} disordered systems, thereby generalizing recent theoretical~\cite{titovfyo}
and numerical~\cite{terraguarnieri} studies in 1D models
of mesoscopic transport. 
We assert that the algebraic decay $P (\Gamma) \sim \Gamma^{-1}$ represents a universal property of Anderson localization of light.  

Although the statistical properties of resonance widths in open
systems have been extensively studied over the last years, in particular for
chaotic/ballistic systems~\cite{distgamma,fyosomm,sommfyotit},
their behavior for systems exhibiting localization has 
received considerably less attention.
As argued by Casati {\it et al.},
$P(\Gamma)$ should follow a power law $P(\Gamma) \sim \Gamma^{-1}$
in localized, classically chaotic systems ~\cite{dima}. 
$P(\Gamma)$ was analytically obtained for 1D disordered systems,
showing a slightly different power law $P(\Gamma) \sim \Gamma^{-1.25}$~\cite{titovfyo}.
This prediction was corroborated later by numerical calculations in
1D and quasi-1D tight binding models~\cite{terraguarnieri}.
The $P (\Gamma) \sim \Gamma^{-1}$ behavior was also reported
in 1D and 3D strongly
driven atomic Rydberg states in the context of dynamical localization~\cite{andreas}. 
Exactly at the Anderson transition~\cite{distcritical} and in the diffusive regime~\cite{distdifusive}, $P(\Gamma)$
was shown to follow a power law with a power 
different from $-1$. Concerning the study of $P (\Gamma)$
for optical systems, the only work on the subject is, to the best of our knowledge,
due to Patra~\cite{patra}, who mainly focused on the small $\Gamma$ regime and its application to random lasers. For small $\Gamma$, it is known that $P (\Gamma)$ is different from a power law, for both the diffusive~\cite{fyosomm,patra} and the localized regimes~\cite{sommfyotit,patra}. It should be emphasized that the power law decay of $P (\Gamma)$ is expected to occur only for intermediate values of $\Gamma$, i.e., for not too large $\Gamma$, where the resonances are strongly coupled to the continuum and where $P (\Gamma)$ decays faster than algebraically~\cite{titovfyo,terraguarnieri}, and for $\Gamma \gtrsim  \langle \Gamma \rangle$.

We will present a simple physical argument,
inspired by Refs.~\cite{dima,titovfyo,terraguarnieri},
to explain the {\em universal} $P (\Gamma) \sim \Gamma^{-1}$ behavior for the localized regime,
i.e., independent on the dimensionality of the system. Due to the opening of the system, exponentially localized eigenstates of the corresponding closed system (linear size $R$)  
acquire a finite frequency width $\Gamma^{\prime}$,
$\Gamma^{\prime} \sim e^{- r^{\prime}/ \xi}$, with $r^{\prime}$ the distance to the boundaries.
Near the system edges, the leakage is strong and the 
resonances are large compared to $\Gamma_{T}$. On the other hand,
far from the boundaries the leakage is
small and the typical $\Gamma$ in this region is much smaller than $\Gamma_{T}$.
Assuming that the resonances are uniformly 
distributed in space, it follows that the (integrated) probability
of finding a resonance width $\Gamma$ smaller than $\Gamma^{\prime}$, 
$P_{int} \left( \Gamma  <  \Gamma^{\prime}  \right)$, is equal
to the probability of finding a resonance situated at a
distance $r$ from the boundaries larger than $r^{\prime}$, $P \left( r >  r^{\prime}  \right)$, i.e.,
$P_{int} \left( \Gamma  <  \Gamma^{\prime}  \right) = P \left( r >  r^{\prime}  \right)$.  
Since $P \left( r >  r^{\prime}  \right) \propto \mu_{D} \left( R - r^{\prime} \right)/
\mu_{D} \left( R \right)$ with $\mu_{D}$ the D-dimensional volume, we conclude that
the probability density is
\begin{eqnarray} 
P \left( \Gamma^{\prime} \right) & = & \frac{d r^{\prime}}{d \Gamma^{\prime}} \frac{d}{dr^{\prime}}
\left[P (r > r^{\prime})\right]  \nonumber \\   
& \propto & - \frac{\xi}{\Gamma^{\prime}} \frac{d}{d r^{\prime}} \left[ \frac{\mu_{D} \left( R - r^{\prime} \right)}
{\mu_{D} \left( R \right)} \right].
\label{probfinal}
\end{eqnarray}
The purely geometrical
factor $ d / d r^{\prime} \left[ \mu_{D} \left( R - r^{\prime} \right)/
\mu_{D} \left( R \right) \right]$ depends on the dimensionality
of the system but does not affect the exponent in $\Gamma$.

To test the validity of (\ref{probfinal})
in the context of Anderson localization of light, we
will consider scalar wave propagation in disordered media
using the method introduced
by Rusek and Orlowski~\cite{ro,ro00}.
This approach is based on the analysis 
of the Green matrices spectra, which describe light scattering from 
randomly distributed pointlike dipoles
(i.e., particles much smaller than the wavelength of light).
For an incident plane wave $\psi_{0} \left( {\mathbf r} \right)$
in a system of $N$ identical dipoles with scattering matrix $t$,
the field acting in the dipole at $\mathbf{r}_{i}$
is given by~\cite{ro,ro00}:
\begin{equation}
\psi \left( \mathbf{r}_{i} \right) = \psi_{0} \left( {\mathbf r}_{i} \right)
+ t \sum\limits_{j \neq i}^{N} G \left( {\mathbf r}_{ij} \right) 
\psi \left( {\mathbf r}_{j} \right).
\label{multiplescatt}
\end{equation} 
The complex-valued $N \times N$ matrix $G \left( {\mathbf r}_{ij} \right)$ describes
light propagation of the wave scattered
by the dipole at ${\mathbf r}_{i}$ to the dipole at ${\mathbf r}_{j}$.
Since the eigenvalues $\lambda_{M}$ of ${\mathbf M} \equiv {\mathbf I} - t{\mathbf G}$ and $\lambda_{G}$ of ${\mathbf G}$
are related by $\lambda_{M} = 1 - t \lambda_{G}$, and $t$ depends on frequency $\omega$ via the scattering phase shift $\delta \left( \omega \right)$~\cite{green}, an eigenvalue $\lambda_{G}$
with $ \text{Re} \lambda_{G}= -1$ will permit an appropriate choice for 
$\delta \left( \omega \right)$ in order to obtain an eigenvalue $\lambda_{M} = 0$, which would correspond to a genuinely localized state somewhere inside the random medium~\cite{ro}.
Assuming a Breit-Wigner model for the scatterers,
(with one resonance
of width $\Gamma_{0}$ at the position $\omega_{0}$) and for which $\delta \left( \omega \right)$ has a simple form, it is possible to obtain, in a good approximation, the resonance widths $\Gamma$ via $\lambda_{G}$,
$\Gamma / \Gamma_{0} \simeq 1 + \text{Re} \lambda_{G}$~\cite{ro00}. 
In the following, we will numerically diagonalize ${\mathbf G}$ in 1D, 2D and 3D
and calculate the distribution of resonance widths
$P \left(  \Gamma  \right)$ using the above approximation.

In Fig.~(\ref{1D}), $P \left(  \Gamma  \right)$ is calculated for 1D systems composed of $100$ randomly
distributed scatterers in a linear segment for two different values of the uniform
optical density $\rho$: $\rho = 1$ and $\rho = 10$ scatterers per wavelength. 
In 1D, all eigenstates are known to be exponentially localized
even for weak disorder and $\xi$ is of the order of the mean free path $\ell$.
$P \left(  \Gamma  \right)$ exhibits
a power law with an exponent
very close to $-1$, in good agreement with (\ref{probfinal}).
In addition, the exponent does not change by increasing 
$\rho$, i.e., by decreasing $\xi$.
This demonstrates that the algebraic decay $P \left(  \Gamma  \right) \sim \Gamma^{-1}$
in the localized regime is valid not only for 1D models of mesoscopic transport~\cite{titovfyo,terraguarnieri}, but also for 
our model of wave propagation in disordered media.
At large $\Gamma$, $P \left(  \Gamma  \right)$ decays
faster than algebraically. This can be explained
by the fact that this region is dominated by short living
resonances, typically close to the boundaries, for which the prediction (\ref{probfinal}) breaks down.

Fig.~(\ref{2D}) shows $P \left(  \Gamma  \right)$ for
2D systems of $2500$ scatterers randomly
distributed in a square for $\rho = 1$ and $\rho = 10$
scatterers per wavelength squared.
According to the scaling theory of localization,
all eigenstates in 2D closed systems should be localized, but
$\xi$ scales exponentially with $\ell$~\cite{costas}.
The $\Gamma^{-1}$ decay of $P \left(  \Gamma  \right)$ in Fig.~(\ref{2D})
is clearly visible for both values of $\rho$ used,
with an exponent very close
to $-1$, in excellent
agreement with (\ref{probfinal}).
Remark however that the range of the power law is broader when $\rho$ is higher. Increasing
$\rho$ means decreasing $\ell$ and, consequently,
decreasing $\xi$. The range of the algebraic decay $P \left(  \Gamma  \right) \sim \Gamma^{-1}$
is expected to be broader as more and more states become localized. 
Such a behavior
was also reported in numerical calculations within
the Anderson model~\cite{terraguarnieri}.
For large $\Gamma$, $P \left(  \Gamma  \right)$ decays again faster than algebraically as in the 1D case.

In Fig.~(\ref{3D}) the 3D case is considered, where 
$P \left(  \Gamma  \right)$ is calculated for
systems composed by $1000$ point scatterers randomly
distributed in a sphere (radius $R$) for $\rho = 1$, $\rho = 10$,
$\rho = 30$ and $\rho = 60$ scatterers per wavelength cubed. 
In 3D, the system
is expected to undergo, upon varying the degree of disorder,
a transition from extended states to localized states. It is therefore
interesting to investigate if and how this transition manifests itself in $P \left(  \Gamma  \right)$.
As in the 2D case, we notice that, as $\rho$ increases, the range of the algebraic decay $P \left(  \Gamma  \right) \sim \Gamma^{-\alpha}$ increases. 
We also remark that, as $\rho$ increases, the {\em associated exponents tend more and more to the value $-1$}. The exponents, obtained by a linear fit in the range where the power law is present, are $\alpha \approx 0.76$ for $\rho = 1$, $\alpha \approx 0.83$ for $\rho = 10$, $\alpha \approx 0.95$ for $\rho = 30$ and $\alpha \approx 1.1$ for $\rho = 60$. This suggests, according to (\ref{probfinal}), the onset of the localized regime for higher $\rho$. In fact, the Ioffe-Regel criterion for localization ($k \ell < 1$) is estimated to be satisfied for $\rho > 2 \pi^{2} \approx 20$ for scatterers at resonance.  
This condition is fulfilled for $\rho = 30$ and $\rho = 60$, for which $\alpha$ is very close to $1$, showing that the system with these densities are indeed in the localized regime and
confirming that the power law $P \left(  \Gamma  \right) \sim \Gamma^{-1}$ can be considered a genuine signature of Anderson localization of light. The fact that the exponents for $\rho = 30$ and $\rho = 60$ are not exactly equal to $1$ can probably be attributed to finite-size effects.
Once again, note that $P \left(  \Gamma  \right)$ decays faster than a power law for large $\Gamma$.  

Let us now compare the mean width of the distribution 
to the inverse Thouless time $\Gamma_{T} = 1/ t_{T} = 6 D_{B} / R^{2}$,
where $D_{B}$ is the Boltzmann diffusion constant. To
estimate $\Gamma_{T}$ one should recall that in 3D,  
$D_{B} = v_{E} \ell^{*}/3$, where $\ell^{*}$ is the
transport mean free path (which is, for point scatterers,
equal to $\ell$) and
$v_{E}$ the energy transport velocity,
$v_{E} \approx c_{0} / [1 + \tau_{dwell}/ \tau_{mf}]$~\cite{reports}, with
$\tau_{dwell} = 1/ \Gamma_{0}$ the dwell time in a single scattering and $\tau_{mf} = \ell/c_{0}$
the mean free time. 
Applying these considerations, $\Gamma_{T}$
can be written as
$\Gamma_{T} / \Gamma_{0} \sim 2 \left( \ell / R \right)^{2}$. 
Fig.~(\ref{3Dthouless}) exhibits $P \left(  \Gamma  \right)$ for the same optical densities of Fig.~(\ref{3D}) but now with $\Gamma$ normalized to $\Gamma_{T}$. For low $\rho$ ($\rho = 1$ and $\rho = 10$), $P \left(  \Gamma  \right)$ is peaked near $\Gamma_{T}$, showing that the system is in the diffusive regime. Notice that there is a non-vanishing probability to find modes that live much longer than $t_{T}$ even in the diffusive regime, the so-called ``prelocalized" modes~\cite{preloc}. 
As $\rho$ increases, we observe that $P \left(  \Gamma  \right)$ is no longer centered at $\Gamma_{T}$ and that the probability to find a mode with resonance width smaller than $\Gamma_{T}$ also increases. This means that, on average, the modes live longer than $t_{T}$.
At the same time, Fig.~(\ref{3D}) shows that localization manifests itself in $P \left(  \Gamma  \right)$ not only via the broadening of the power law range but also via the fact that the associated exponents approach to $-1$. 
We conclude again that the $P \left(  \Gamma  \right) \sim \Gamma^{-1}$ behavior is an unambiguous signature of Anderson localization of light in open media.
It must be mentioned that the present 3D study may be relevant for recent multiple light scattering experiments in atomic media~\cite{nice}, for which modeling the scatterers by pointlike dipoles constitutes an excellent approximation.

In summary, we have studied the distribution of resonance widths $P \left(  \Gamma  \right)$
in 1D, 2D and 3D multiple light scattering systems composed of randomly distributed pointlike scalar dipoles. We have developed a simple physical argument, based on the exponential decay of localized eigenfunctions, to show that $P \left(  \Gamma  \right)$ should follow an universal power law $P \left(  \Gamma  \right) \sim \Gamma^{-1}$ decay in the localized regime. This prediction was confirmed by extensive numerical calculations and demonstrates that the $P \left(  \Gamma  \right) \sim \Gamma^{-1}$ behavior can be interpreted as an unambiguous signature of Anderson localization of light in open media.
    
%
%
\acknowledgments
	
Most of the computations presented in this paper were performed on 
the cluster PHYNUM (CIMENT, Grenoble). This work was supported 
in part by the Polonium contract 03290 VK. M.R. and A.O. were supported by 
the Polish State Committee for Scientific Research (KBN) under Grant 
no. 2 P03B 044 19. F.A.P. 
wishes to thank CNPq/Brazil for financial support.

%
%
%
%

%
%
%
%
%
%
\newpage

\begin{figure}
\caption{The normalized distribution of resonance widths $P \left(  \Gamma  \right)$ calculated for
$1000$ different configurations of $100$ point scatterers randomly distributed
in a 1D segment with two different values of the uniform optical density $\rho$,
$\rho = 1$ (full squares) and $\rho = 10$ (open circles) scatterers per wavelength. The
dashed line corresponds to the prediction $P \left(  \Gamma  \right) \sim \Gamma^{-1}$
for the localized regime and the solid ones are just to guide the eyes. The values of $\Gamma$
are normalized by the resonance width of a single dipole $\Gamma_{0}$.} \label{1D}
\end{figure} 

\begin{figure}
\caption{$P \left(  \Gamma  \right)$ calculated for up to
$50$ configurations of $2500$ scatterers randomly distributed
in a square for $\rho = 1$ (full squares) and $\rho = 10$ (open circles) scatterers per wavelength squared. The normalization of $\Gamma$, as well as the significance of the solid and dashed lines, is the same as in Fig.~(\ref{1D}).} \label{2D}
\end{figure} 

\begin{figure}
\caption{$P \left(  \Gamma  \right)$ calculated for 
$100$ configurations of $1000$ scatterers randomly distributed
in a sphere for
$\rho = 1$ (full squares), $\rho = 10$ (open circles), $\rho = 30$ (full triangles) and $\rho = 60$ (open diamonds) scatterers per wavelength cubed.
The normalization of $\Gamma$, as well as the significance of the solid and dashed lines, is the same of Fig.~(\ref{1D}).} \label{3D}
\end{figure} 

\begin{figure}
\caption{$P \left(  \Gamma  \right)$ as in Fig.~(\ref{3D}), but now $\Gamma$ is normalized by the Thouless frequency, $\Gamma_T$.} \label{3Dthouless}
\end{figure} 


\begin{thebibliography}{99}
\bibitem{green} The $t (\omega)$ matrix relates to the phase shift $\delta (\omega)$
according to $t (\omega) = a \left\{ \exp \left[2i \delta (\omega) \right] - 1 \right\}$, with $a$ a real-valued constant. Actually, ${\mathbf G}$ is the dimensionless Green matrix in D dimensions (see Ref.~\cite{ro}).
\end{thebibliography}
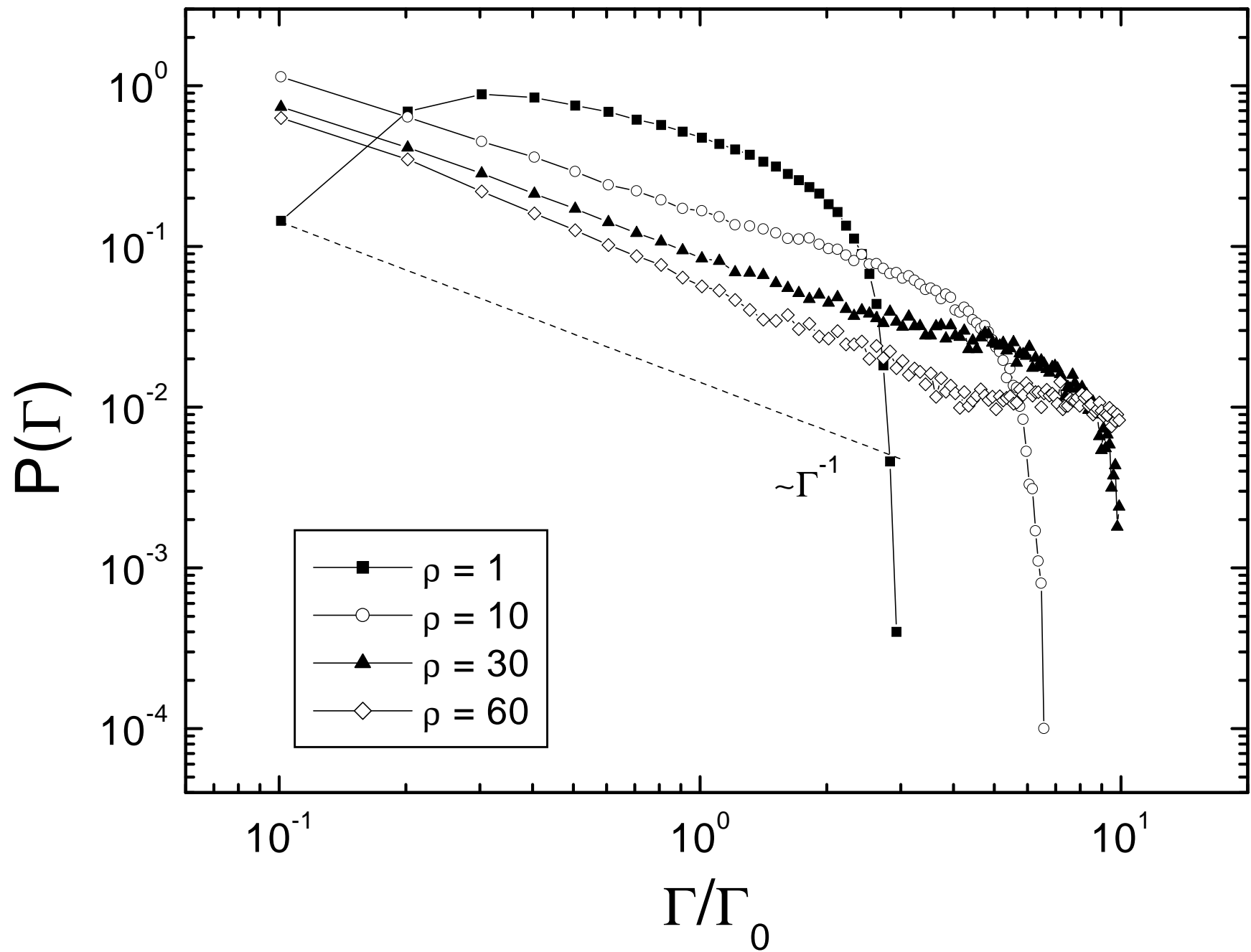
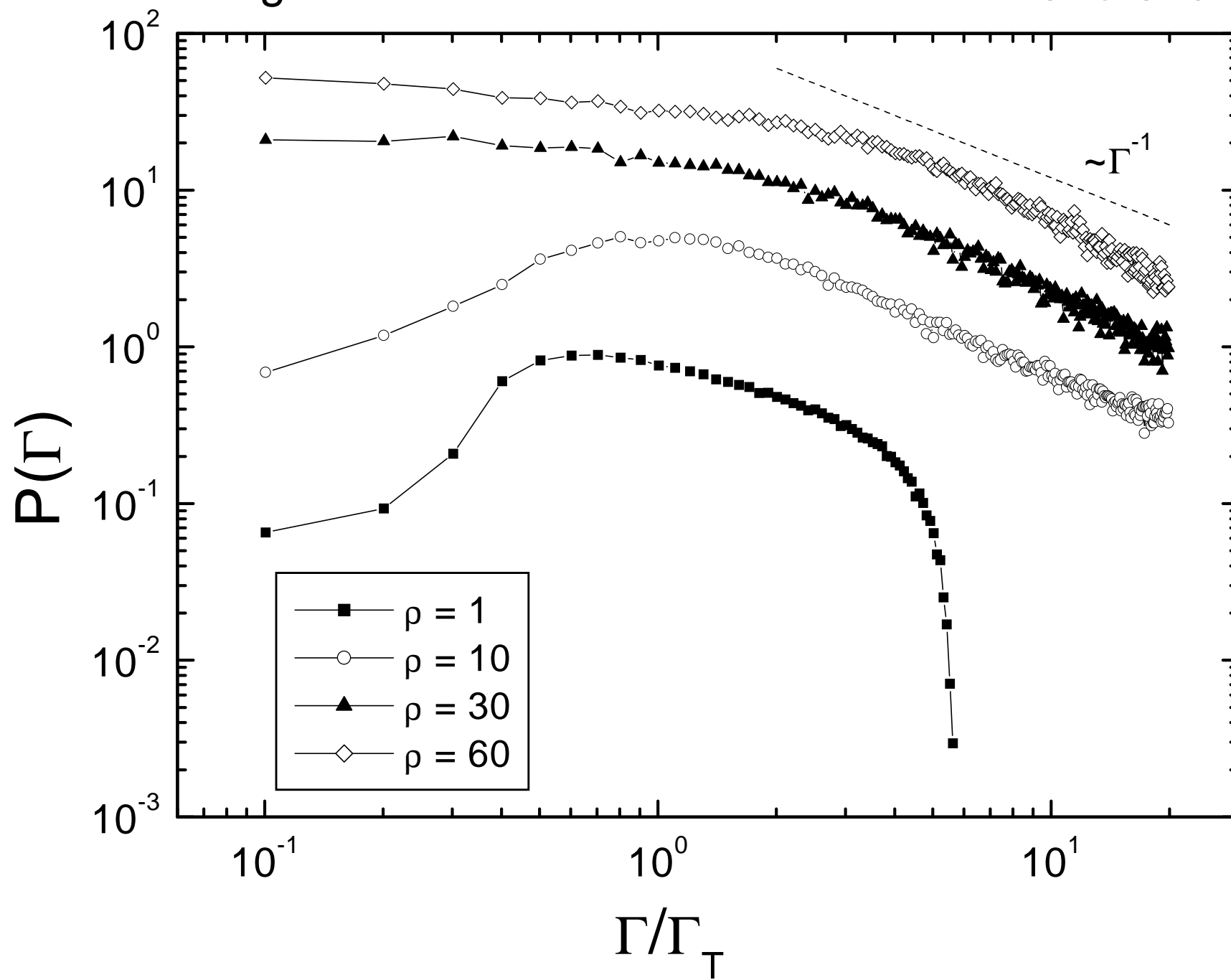
\end{document}